\documentclass[apj]{emulateapj}
\usepackage{apjfonts}
\begin{document}
\newcommand{\frip}{Fish \& Reid, in preparation{}}
\shorttitle{Magnetic Fields and Maser Motions in W75~N}
\title{Large Magnetic Fields and Motions of OH Masers in W75~N}
\shortauthors{Fish \& Reid}
\author{Vincent L.\ Fish\altaffilmark{1} \& Mark J.\ Reid\altaffilmark{2}}
\altaffiltext{1}{Jansky Fellow, National Radio Astronomy Observatory,
  1003 Lopezville Road, Socorro, NM 87801}
\email{vfish@nrao.edu}
\altaffiltext{2}{Harvard--Smithsonian Center for Astrophysics, 60
  Garden St., MS 42, Cambridge, MA 02138}
\begin{abstract}
We report on a second epoch of VLBA observations of the 1665 and
1667~MHz OH masers in the massive star-forming region W75~N.  We find
evidence to confirm the existence of very strong ($\sim 40$~mG)
magnetic fields near source VLA~2.  The masers near VLA~2 are
dynamically distinct and include a very bright spot apparently moving
at $50$~km\,s$^{-1}$ relative to those around VLA~1.  This fast-moving
spot may be an example of a rare class of OH masers seen in outflows
in star-forming regions.  Due to the variability of these masers and
the rapidity of their motions, tracking these motions will require
multiple observations over a significantly shorter time baseline than
obtained here.  Proper motions of the masers near VLA~1 are more
suggestive of streaming along magnetized shocks rather than Keplerian
rotation in a disk.  The motions of the easternmost cluster of masers
in W75~N~(B) may be tracing slow expansion around an unseen exciting
source.
\end{abstract}

\section{Introduction}
\label{introduction}

\object[W75N]{W75~N} is a complex massive star-forming region with
many sites of star formation and masers \citep[e.g.,][]{persi06}.
W75~N~(B) is composed of at least four continuum sources, of which
VLA~1 and VLA~2 host both OH and H$_2$O masers
\citep{hunter94,torrelles97}.  The H$_2$O masers around VLA~2 appear
to be located on a shell with an expansion velocity of $\sim
28$~km~s$^{-1}$ \citep{torrelles03} and organized substructure on
scales as small as $1$~AU \citep{uscanga05}.  The sources VLA~1,
VLA~2, and VLA~3 are all believed to drive large-scale molecular
outflows \citep[][and references therein]{shepherd03,shepherd04}.

The OH masers in W75~N were first observed interferometrically by
\citet{harvey74}, who noted three different sites of OH maser
emission.  VLBI techniques were first employed in this source by
\citet{haschick81}.  The full large-scale distribution of masers,
including the easternmost cluster and the group near VLA~2, was
detected by \citet{baart86}.  Extreme variability of OH maser features
in W75~N was noted by \citet{alakoz05}, who reported that a flare up
to $750$~Jy was briefly the brightest OH maser in the sky.  Flaring is
also seen in many of the water \citep[][and references
therein]{lekht00} and methanol masers in this source
\citep{goedhart04}.  The OH masers near VLA~2 appear to have a much
larger velocity scatter than do the rest of the masers in W75~N~(B)
\citep{fish05}.  It is in the aim of exploring the kinematic
differences between the masers in VLA~1 and VLA~2 that we revisit the
region with VLBA observations.

\section{Observations}
\label{observations}

The VLBA was used to observe the 1665.4018 and 1667.3590~MHz
transitions of OH in W75~N (G81.871$+$0.781) on 2000 November 22 and
2001 January 06 (data combined; hereafter, epoch 1).  For each
frequency, dual-circular polarization observations were taken using a
125 kHz bandwidth divided into 128 spectral channels (corresponding to
a velocity width of 0.176~km~s$^{-1}$), centered on v$_\mathrm{LSR} =
10$~km~s$^{-1}$ (radio).  Further details of these observations as
well as the resulting images can be found in \citet{fish05}.  A second
epoch of data was taken on 2004 September 16 and 19.  Observational
parameters were identical, with the exception that the bandwidth was
centered at v$_\mathrm{LSR} = 7$~km~s$^{-1}$.  The synthesized beam
size was approximately $9.0 \times 6.5$~mas with a blank sky rms noise
of 8 mJy~beam$^{-1}$.  In this work we report both on the second epoch
of observations and observed changes in W75~N between the two epochs.

\section{Results}
\label{results}
\begin{figure}[ht]
\resizebox{\hsize}{!}{\includegraphics{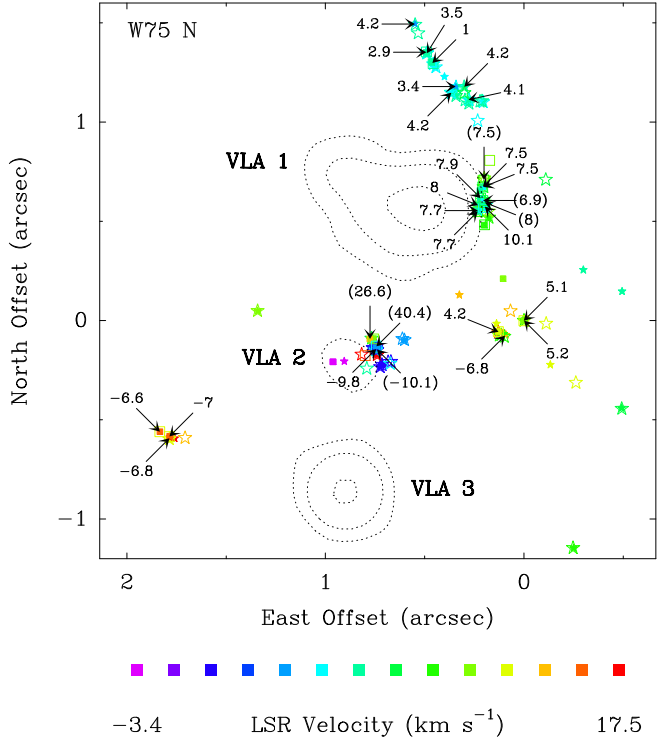}}
\caption{Map of detected OH maser emission in W75~N (epoch 2004 Sep
  16/19).  Stars indicate 1665~MHz emission, and squares indicate
  1667~MHz emission.  Open (filled) symbols denote emission in RCP
  (LCP).  The LSR velocity of the peak of each maser profile is
  indicated in color.  Numbers indicate magnetic fields derived from
  Zeeman splitting, with positive values representing fields oriented
  in the hemisphere away from the observer.  Values in parentheses are
  deduced from overlaps in different transitions (see
  Tab.~\ref{brothers-table}).  X-band (8.4 GHz) continuum emission is
  shown in dotted contours.  The feature at the origin is the same as
  the reference feature of \citet{slysh06}, who give its absolute
  position as $20^\mathrm{h}38^\mathrm{m}36\fs416,
  +42^\circ37^\prime34\farcs42 (\pm0\farcs01)$ (J2000, epoch 2001 Jan
  01).
  \label{w75n-vel} }
\end{figure}

\begin{figure}[ht]
\resizebox{\hsize}{!}{\includegraphics{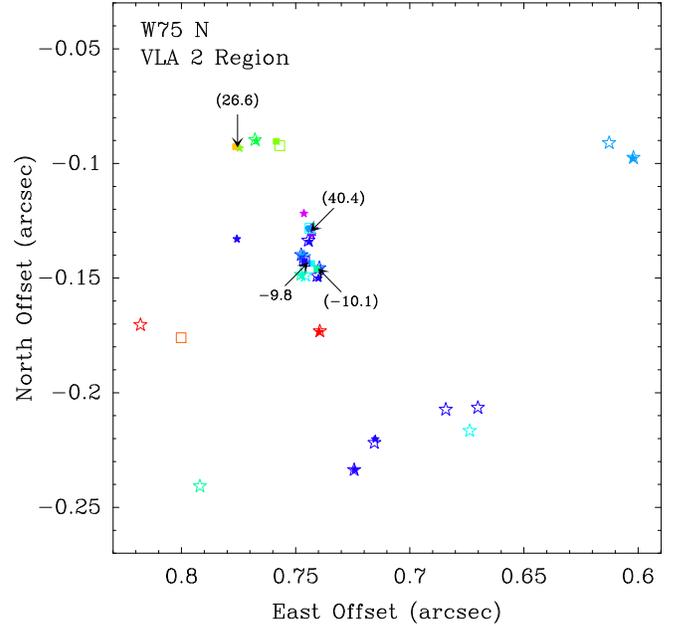}}
\caption{Enlargement of region near VLA~2.  See
  Figure~\ref{w75n-vel} for details.  The continuum contours have been
  suppressed.\label{w75n-vla2}}
\end{figure}

We recover a total of 123 maser features (lines), approximately the
same as the 120 features detected in epoch 1 in the same amount of
on-source observing time \citep{fish05}.  Spot parameters are listed
in Table \ref{maser-table}.  A map of detected maser emission is shown
in Figure \ref{w75n-vel}.  Our velocity coverage includes most of the
brightest features but is not sufficient to have seen all known maser
features in this source, which range from $-4.9$~km\,s$^{-1}$
\citep{hutawarakorn02} up to $+33$~km~s$^{-1}$ \citep{yngvesson75}.
The map is qualitatively similar to epoch 1 \citep[Fig.~27
of][]{fish05} except near VLA~2 (Fig.~\ref{w75n-vla2} of the present
work).

The magnetic fields we infer from Zeeman splitting of left- and
right-circular polarization (LCP and RCP) components are shown in
Figure~\ref{w75n-vel} and listed in Table \ref{zeeman-table}.  The
strength and line-of-sight direction of the magnetic fields are
consistent with values obtained in the epoch 1, except near source
VLA~2.  Our velocity coverage is too small to detect both Zeeman
components at 1665~MHz (and at 1667~MHz, depending on the systemic
velocity) when split by $\sim 40$~mG, as seen by \citet{slysh06} near
VLA~2.  However, we can also infer magnetic field strengths where 1665
and 1667~MHz masers overlap, since the transitions undergo different
Zeeman splitting for a given magnetic field strength.  These values
are presented in parentheses in Figure \ref{w75n-vel} and listed in
Table \ref{brothers-table}.  Near VLA~2 we find magnetic fields of
$26.6$ and $40.4$~mG (Fig.~\ref{w75n-vla2}), consistent in sign and
(at least in the latter case) magnitude with values obtained by
\citet{slysh06}.  (We note that the random error in the magnetic field
strength estimate from multi-transition overlap may be somewhat larger
than from traditional Zeeman pairs, since it is not known a priori
that the 1665 and 1667~MHz Zeeman patterns are centered at the same
systemic velocity.  Nevertheless, this technique generally results in
magnetic field estimates consistent with values obtained from proper
Zeeman pairs, as can be seen near VLA~1 as well as in other sources
(\frip).)

Reanalyzing the data from epoch 1, we agree with \citet{slysh06} that
spots 109 and 110 in Table 15 of \citet{fish05} form a $42.5$~mG
Zeeman pair at 1667~MHz (with multi-transition confirmation from spot
67 at 1665~MHz).  Spots 112 and 113 also form a $34.5$~mG Zeeman pair
(approximately consistent with v$_\mathrm{LSR}$ of nearby 1665~MHz LCP
spots 69 and 70).  However, not all overlaps are indicative of
magnetic fields.  It is unlikely that spots 68 and 107 arise from a
$137$~mG magnetic field, as this would imply a systemic velocity
v$_\mathrm{LSR} = 40.5$~km\,s$^{-1}$, well above the maximum of
detected masers in the region \citep{yngvesson75}.

\begin{figure}[t]
\resizebox{\hsize}{!}{\includegraphics{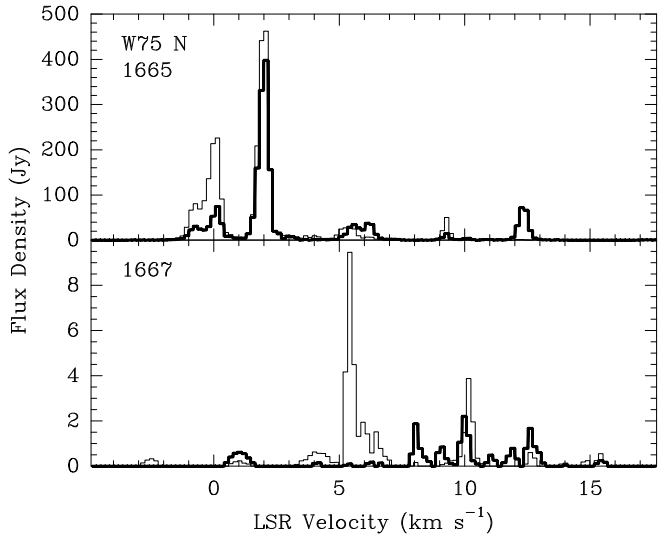}}
\caption{Spectrum of recovered maser emission in W75~N on 2004
  September 16/19.  RCP emission is shown in bold and LCP emission in
  normal weight.
  \label{w75n-spec} }
\end{figure}

The spectrum of recovered maser emission is shown in Figure
\ref{w75n-spec}.  In each velocity channel, the total flux density of
each (elliptical Gaussian) fitted maser spot is summed.  The strongest
maser we detect is located at $(\Delta\alpha,\Delta\delta) =
(739.60,-145.53)$ and is approximately $400$~Jy in both RCP and LCP
($2kT/A$).  The equal flux densities suggest that the maser may be
nearly 100\% linearly polarized, as noted by \citet{alakoz05}.  Other
masers near VLA~2 also show very high linear polarization fractions
\citep{baart86,slysh01,hutawarakorn02,fish05}.  The strongest spot is
over twice as strong as in epoch 1 \citep{fish05} and is stronger than
measured on 2004 October 20 by \citet{alakoz05}.  It is unclear
whether the same maser spot is experiencing a second flare or a
different, nearby maser spot is undergoing a flare episode.

\begin{figure}[t]
\resizebox{\hsize}{!}{\includegraphics{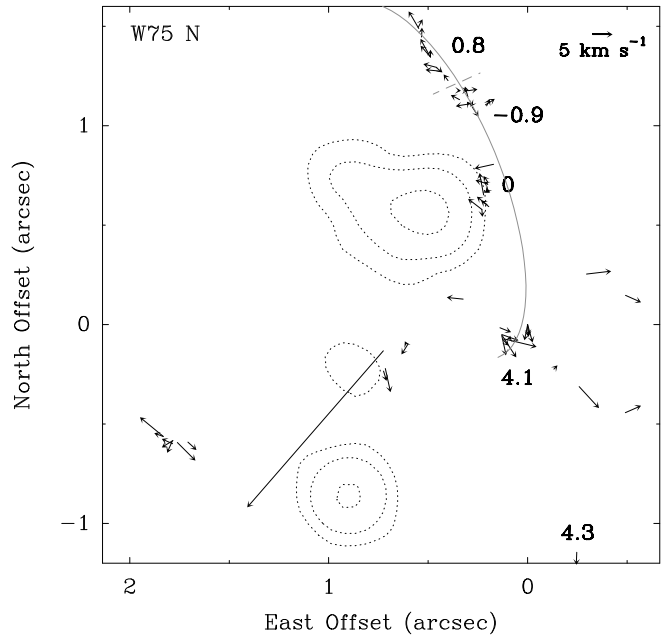}}
\caption{Proper motions of OH masers.  Arrow lengths arge proportional
  to velocity as indicated in the upper right, assuming a distance of
  $2.0$~kpc.  The very long arrow corresponds to the motion of the
  $400$~Jy maser near VLA~2.  The reference frame has been chosen to
  minimize the total motion of masers not associated with VLA~2.
  Numbers indicate proper motions in km\,s$^{-1}$ with respect to the
  cluster labelled ``0'' (positive indicates expansion).  The gray
  curve shows the approximate location of the disk from
  \citet{slysh01}.
  \label{w75n-motions} }
\end{figure}

Proper motions of the OH masers are shown in Figure
\ref{w75n-motions}.  Since the data for both epochs were phase
referenced to a maser spot rather than an external source, our data
cannot distinguish between the map presented and any other frame with
a single constant vector added to all motions.  In Figure
\ref{w75n-motions} we have chosen the frame that minimizes the total
length of all arrows not associated with VLA~2.  The average speed of
all masers excluding those near VLA~2 is $3.4 \pm 2.1$~km~s$^{-1}$.
This speed is comparable to that of OH masers around other young
ultracompact \ion{H}{2} regions, such as W3(OH) and ON~1
\citep[][\frip]{bloemhof92}.  The random error on the motion of an
individual maser spot, given by the uncertainties in determining the
position centroid in each epoch, ranges from negligible to
1.4~km~s$^{-1}$ with a median value of 0.2~km~s$^{-1}$.

The presence of a disk associated with VLA~1 has often been inferred
from the north-south distribution of OH masers
\citep[e.g.,][]{haschick81,slysh01}.  Indeed, the distribution of LSR
velocity increases from north to south along this structure (e.g.,
Fig.~\ref{w75n-vel}).  However, the proper motions of the OH masers
are less clear.  Figure~\ref{w75n-motions} shows the average proper
motions of clusters of masers along and south of the putative disk,
relative to the cluster nearest the continuum emission (velocity
``$0$'').  Positive values indicate net motion away from this cluster.
Proper motions in the south show clear organization away from the
``0'' cluster, but the motions of the northernmost clusters are mixed.
Alternatively, the masers may be tracing a shocked, expanding layer
\citep{baart86}.  If so, the maser motions in the northernmost cluster
may include a streaming component along an organized magnetic field
\citep{fish06}, with the caveat that while the field direction appears
to be oriented along the distribution of masers, the external Faraday
rotation contribution is unknown \citep[see also][]{slysh02}.

The motion of the $400$~Jy maser (feature 70 in Table
\ref{maser-table}) near VLA~2 is over $50$~km~s$^{-1}$.  Other than
this particular maser, it is difficult to conclusively identify masers
from one epoch to the next near VLA~2 due to variability and the large
number of masers in the region.  The distribution of masers near VLA~2
is also changing rapidly, as can be seen from comparison of
Figure~\ref{w75n-vla2} with Figure~28 of \citet{fish05}.  LSR velocity
is not necessarily a reliable indicator, as the masers appear to be
accelerating.  The brightest maser in the region has seen a change in
velocity $\Delta \mathrm{v}_\mathrm{LSR} = -0.45$~km~s$^{-1}$ between
the two epochs.  \citet{alakoz05} note a similar decrease in
v$_\mathrm{LSR}$ of this feature over a 2.5-year timescale.

The easternmost cluster of OH masers may be associated with another
exciting source not seen in the continuum maps.  The velocity of this
cluster corrected for Zeeman splitting is $\sim 14$~km~s$^{-1}$,
higher than any other cluster in W75~N (except for portions of VLA~2,
which has a huge velocity scatter and no clear organization).
Additionally, the maser proper motions are suggestive of slow ($\sim
5$~km~s$^{-1}$) expansion from a point in the center of the cluster
(Fig.~\ref{w75n-motions}).  This motion is likely preferentially in
the plane of the sky in light of the small LSR velocity scatter of the
masers.

\section{Discussion}
\label{discussion}

It is difficult to track OH motions from the first epoch to the second
epoch near VLA~2, both due to variability and the fast motions
involved.  These properties may be interrelated.  It is probably not a
coincidence that the two strongest ($> 100$~Jy) masers in W75~N
(features 70 and 73) both appear in the complicated cluster around
VLA~2.  While a connection between kinetic energy and H$_2$O maser
flux density has been clearly established \citep{anglada96}, such a
correlation is far less clear for OH, which is believed to be
predominantly radiatively pumped.  Nevertheless, regions with
energetic motions may change the balance of pumping routes in favor of
those with collisional components, possibly increasing the net pump
rates in certain ground-state lines.  (For a fuller discussion in a
different context, see \citet{gray05}.)  We speculate that the energy
source powering the VLA~2 outflow also helps to produce one of the
strongest OH masers in the sky.

In most OH maser sources, it is fairly well established that the
observed proper motions of the masers correspond to motions of
physical clumps of material (see \citealt{bloemhof96} and \frip).  But
it is perhaps possible that the apparent motion of the brightest maser
in VLA~2 is due to the passage of a fast-moving shock front through
different portions of a cloud of OH rather than the ballistic motion
of a single high-velocity clump of material.  In any event, it appears
that most individual maser spots near VLA~2 do not persist over
timescales of years or longer.  Multi-epoch phase-referenced VLBI
observations spaced several months apart will be essential to track
the motions of the OH masers around VLA~2.  Phase referencing to a
nearby bright calibrator would eliminate the ambiguity of the absolute
position of the reference spot in each epoch.  Despite this ambiguity
in our data, it is clear that the $50$~km\,s$^{-1}$ motion of feature
70 in Table \ref{maser-table} with respect to the other masers in
W75~N is due to a large intrinsic motion of feature 70 itself, since
the velocity scatter (both from proper motions and radial velocities)
of the rest of the W75~N masers is small.

\citet{argon03} report on a class of OH masers associated with
outflows.  For an OH maser in the Turner-Welch (TW) object near
W3(OH), they obtain a (projected) proper motion of $66$~km~s$^{-1}$
with $\Delta \mathrm{v}_\mathrm{LSR} = 0.5$~km~s$^{-1}$ over a period
of 8 years.  This is remarkably similar to what is seen near VLA~2.  A
key difference, however, is that the water masers in the TW object
span a range in v$_\mathrm{LSR}$ of about $100$~km~s$^{-1}$
(\citealt{cohen79}; Hachisuka et al., in preparation) with proper
motions of several tens of km~s$^{-1}$.  In comparison, the water
masers in VLA~2 span a range of about $25$~km~s$^{-1}$, are arranged
in a fairly organized circular pattern, and appear to be expanding at
$\sim 28$~km~s$^{-1}$ \citep{torrelles03}, slower than the OH masers.

The magnetic field near VLA~2 appears to undergo a line-of-sight
reversal across this very compact source.  (We note the caveat,
however, that three of the four magnetic field measurements are
inferred from multi-transition overlap rather than single-transition
Zeeman splitting.)  Given the rapidly changing nature of this source,
it would be fruitful to attempt confirmation of the magnetic field at
another epoch, especially since new masers that appear may provide
magnetic field information in regions presently unprobed by Zeeman
splitting.  The polarity of the field may be important if, for
instance, the masers trace a magnetically-collimated outflow
\citep{cunningham05}.  While the exact relation between magnetic field
strength ($B$) and density ($\rho$) may depend on the particular model
\citep[e.g.,][]{pudritz06}, it is possible that $40$~mG magnetic
fields do not necessarily imply the extremely large densities that
would be obtained from collapse models, where $B \propto \rho^\kappa,
\kappa \approx 0.5$ \citep{crutcher99}.  Possibly these masers are
from material being ejected from a protostellar disk, which may
involve strong magnetic fields originating in the disk.  Future
observations may also be able to determine whether the unusually
strong magnetic field at the OH masers in VLA~2 is decaying as in
Cep~A \citep{cohen90,bartkiewicz05}.


\acknowledgments

  The National Radio Astronomy Observatory is a facility of the
  National Science Foundation operated under cooperative agreement by
  Associated Universities, Inc.

{\it Facility: \facility{VLBA}}

\LongTables
\begin{deluxetable}{lrrrrrl}
\tabletypesize{\small}
\tablecaption{Detected Maser Spots\label{maser-table}}
\tablehead{
  \colhead{Freq.} &
  \colhead{Maser} &
  \colhead{$\Delta$ RA} &
  \colhead{$\Delta$ Dec} &
  \colhead{v$_\mathrm{LSR}$} &
  \colhead{Brightness} &
  \colhead{} \\
  \colhead{(MHz)} &
  \colhead{Feature} &
  \colhead{(mas)} &
  \colhead{(mas)} &
  \colhead{(km\,s$^{-1}$)} &
  \colhead{(Jy\,beam$^{-1}$)} &
  \colhead{Polarization}
}
\startdata
1665 &   1 &  $-$493.39 &     147.93 &    6.83 &   0.35 & L \\
     &   2 &  $-$491.55 &  $-$440.64 &    7.35 &   0.07 & L \\
     &     &  $-$491.89 &  $-$443.17 &    7.35 &   0.06 & R \\
     &   3 &  $-$298.29 &     256.13 &    5.95 &   1.03 & L \\
     &   4 &  $-$260.68 &  $-$312.38 &   12.74 &   1.22 & R \\ 
     &   5 &  $-$246.81 & $-$1144.76 &    9.84 &   0.25 & R \\
     &     &  $-$249.22 & $-$1145.51 &    9.81 &   0.13 & L \\
     &   6 &  $-$132.16 &  $-$221.87 &   12.10 &   0.06 & L \\
     &   7 &  $-$111.20 &   $-$14.49 &   12.80 &   0.82 & R \\
     &   8 &  $-$109.22 &     710.59 &    8.23 &   0.15 & R \\
     &   9 &    $-$0.24 &    $-$0.17 &    9.27 &  38.71 & L \\
     &     &       0.16 &       0.37 &    9.28 &   2.89 & R \\
     &  10 &       0.00 &       0.00 &   12.36 &  72.01 & R \\
     &  11 &      68.21 &      49.04 &   13.17 &   0.24 & R \\
     &  12 &      91.69 &   $-$70.85 &   13.68 &   0.33 & L \\
     &  13 &      99.89 &   $-$80.27 &    9.64 &   0.15 & R \\
     &  14 &     115.69 &   $-$71.09 &    9.45 &   0.41 & R \\
     &  15 &     124.48 &   $-$74.32 &   10.01 &   0.25 & R \\
     &  16 &     127.59 &   $-$55.61 &   12.09 &   0.25 & L \\
     &  17 &     129.00 &   $-$58.36 &   14.56 &   0.06 & R \\
     &  18 &     129.22 &   $-$48.70 &   13.33 &   0.23 & L \\
     &  19 &     138.83 &   $-$14.49 &   12.80 &   1.25 & L \\
     &  20 &     177.24 &     520.20 &    9.99 &   0.58 & R \\
     &     &     177.24 &     520.20 &    9.99 &   0.27 & L \\
     &  21 &     178.97 &     550.75 &   10.16 &   0.23 & R \\
     &  22 &     193.50 &     579.87 &    9.99 &   0.13 & R \\
     &  23 &     193.54 &     581.00 &    5.24 &   0.14 & R \\
     &  24 &     196.35 &     595.53 &    8.97 &   1.13 & R \\
     &  25 &     200.58 &     718.00 &   11.22 &   0.34 & R \\
     &  26 &     200.94 &     707.45 &   10.87 &   1.11 & R \\
     &  27 &     201.09 &     568.85 &    4.01 &   1.17 & L \\
     &  28 &     203.33 &     673.95 &    4.54 &   0.19 & L \\
     &  29 &     203.33 &     673.95 &    8.96 &   0.68 & R \\
     &  30 &     205.54 &     605.61 &    9.26 &  11.32 & R \\
     &  31 &     206.33 &    1102.14 &    5.49 &   6.55 & R \\
     &     &     205.82 &    1102.49 &    5.42 &   0.91 & L \\
     &  32 &     211.63 &    1106.88 &    5.90 &   3.03 & R \\
     &  33 &     212.98 &     613.36 &    4.56 &   1.37 & L \\
     &  34 &     221.71 &    1104.56 &    5.95 &   0.39 & R \\
     &  35 &     230.48 &     581.26 &    5.41 &   4.10 & L \\
     &     &     230.63 &     581.31 &    5.48 &   0.44 & R \\
     &  36 &     231.83 &     581.66 &   10.15 &   0.43 & R \\
     &  37 &     231.93 &     549.56 &   10.11 &   3.08 & R \\
     &  38 &     233.01 &     558.86 &    5.55 &  11.66 & L \\
     &     &     233.29 &     559.16 &    5.53 &   2.43 & R \\
     &  39 &     233.47 &    1007.83 &    5.24 &   0.26 & R \\
     &  40 &     277.46 &    1114.03 &    6.50 &   0.45 & R \\
     &  41 &     277.89 &    1094.50 &    5.82 &   6.05 & R \\
     &     &     279.11 &    1097.13 &    5.59 &   0.92 & L \\
     &  42 &     282.51 &    1113.27 &    4.07 &   5.63 & L \\
     &     &     278.85 &    1113.22 &    4.22 &   0.28 & R \\
     &  43 &     297.23 &    1111.03 &    6.85 &   0.90 & R \\
     &  44 &     300.27 &    1178.27 &    3.21 &   0.84 & L \\
     &  45 &     300.94 &    1147.54 &    7.35 &   0.25 & R \\
     &  46 &     301.64 &    1178.27 &    5.70 &   4.93 & R \\
     &     &     301.59 &    1178.09 &    5.59 &   3.25 & L \\
     &  47 &     325.79 &     130.03 &   13.90 &   0.41 & L \\
     &  48 &     342.64 &    1177.95 &    5.51 &  15.39 & R \\
     &     &     342.01 &    1178.43 &    5.57 &   1.66 & L \\
     &  49 &     343.12 &    1133.25 &    6.52 &   2.57 & R \\
     &     &     342.91 &    1132.95 &    6.47 &   0.39 & L \\
     &  50 &     346.17 &    1180.21 &    3.48 &   0.38 & L \\
     &  51 &     365.98 &    1147.96 &    4.54 &   0.26 & L \\
     &     &     365.89 &    1148.01 &    4.54 &   0.08 & R \\
     &  52 &     367.61 &    1149.59 &    7.01 &   0.18 & R \\
     &  53 &     400.92 &    1229.55 &    4.71 &   0.56 & L \\
     &  54 &     445.02 &    1276.97 &    5.31 &   1.00 & L \\
     &     &     444.40 &    1277.38 &    5.15 &   0.32 & R \\
     &  55 &     457.52 &    1293.28 &    5.15 &   7.49 & L \\
     &     &     457.14 &    1292.72 &    5.18 &   4.64 & R \\
     &  56 &     488.92 &    1347.51 &    5.16 &   6.15 & L \\
     &     &     488.92 &    1347.51 &    5.07 &   0.18 & R \\
     &  57 &     489.71 &    1348.16 &    7.21 &   0.74 & R \\
     &  58 &     490.24 &    1347.86 &    5.18 &   1.05 & R \\
     &  59 &     533.26 &    1448.99 &    7.00 &   0.16 & R \\
     &  60 &     548.23 &    1491.99 &    6.15 &   0.48 & R \\
     &  61 &     550.47 &    1496.78 &    3.69 &   5.04 & L \\
     &  62 &     602.08 &   $-$97.43 &    3.11 &   2.03 & R \\
     &     &     602.28 &   $-$97.74 &    3.14 &   1.30 & L \\
     &  63 &     612.70 &   $-$91.00 &    3.13 &   0.85 & R \\
     &  64 &     670.20 &  $-$206.46 &    0.50 &   0.51 & R \\
     &  65 &     673.82 &  $-$216.53 &    4.54 &   0.23 & R \\
     &  66 &     684.26 &  $-$207.25 &    0.67 &   0.47 & R \\
     &  67 &     715.54 &  $-$221.75 &    0.67 &   0.43 & R \\
     &     &     715.14 &  $-$220.16 &    0.85 &   0.36 & L \\
     &  68 &     724.38 &  $-$233.62 &    0.60 &   0.34 & R \\
     &     &     724.16 &  $-$233.53 &    0.67 &   0.32 & L \\
     &  69 &     739.42 &  $-$173.13 &   17.40 &   0.13 & R \\
     &     &     739.72 &  $-$173.37 &   17.39 &   0.08 & L \\
     &  70 &     739.60 &  $-$145.53 &    2.00 & 408.83 & R \\
     &     &     739.39 &  $-$145.95 &    1.98 & 402.02 & L \\
     &  71 &     740.07 &  $-$150.02 & $-$0.12 &  51.31 & L \\
     &     &     740.61 &  $-$148.88 & $-$0.13 &   9.32 & R \\
     &  72 &     742.92 &  $-$130.89 & $-$0.77 &  52.19 & L \\
     &     &     743.08 &  $-$128.94 & $-$0.75 &  23.14 & R \\
     &  73 &     744.08 &  $-$134.10 &    0.09 & 134.88 & L \\
     &     &     744.44 &  $-$133.57 &    0.11 &  59.95 & R \\
     &  74 &     745.71 &  $-$148.79 &    4.26 &   0.48 & R \\
     &  75 &     746.35 &  $-$140.72 &    3.09 &   1.25 & R \\
     &  76 &     746.38 &  $-$121.79 & $-$3.37 &   0.73 & L \\
     &  77 &     747.32 &  $-$140.04 &    2.69 &   5.61 & L \\
     &     &     747.57 &  $-$139.85 &    2.43 &   4.58 & R \\
     &  78 &     747.74 &  $-$148.44 &    6.21 &  34.73 & R \\
     &     &     747.34 &  $-$148.90 &    6.23 &   4.89 & L \\
     &  79 &     767.28 &   $-$90.12 &    7.14 &   1.08 & L \\
     &     &     767.82 &   $-$89.66 &    7.23 &   0.32 & R \\
     &  80 &     774.58 &   $-$93.27 &   10.87 &   0.14 & L \\
     &  81 &     775.76 &  $-$132.90 &    0.85 &   0.39 & L \\
     &  82 &     791.97 &  $-$240.71 &    5.83 &   1.72 & R \\
     &  83 &     817.99 &  $-$170.32 &   17.24 &   0.57 & R \\
     &  84 &     905.22 &  $-$204.56 & $-$2.32 &   0.30 & L \\
     &  85 &    1341.64 &      49.55 &   11.39 &   0.14 & R \\
     &     &    1341.26 &      48.92 &   11.39 &   0.11 & L \\
     &  86 &    1706.84 &  $-$590.03 &   13.68 &   0.35 & R \\
     &  87 &    1759.06 &  $-$592.88 &   16.49 &   0.07 & L \\
     &  88 &    1784.06 &  $-$587.59 &   16.14 &   0.07 & L \\
     &  89 &    1785.98 &  $-$596.61 &   12.16 &   6.42 & R \\
     &     &    1785.19 &  $-$596.61 &   12.14 &   0.30 & L \\
1667 &  90 &       1.10 &    $-$0.61 &   11.89 &   0.76 & R \\
     &  91 &       1.39 &    $-$0.06 &   10.09 &   0.39 & L \\
     &  92 &     105.78 &     211.80 &   10.19 &   1.09 & L \\
     &  93 &     129.66 &   $-$54.79 &   12.62 &   0.49 & R \\
     &     &     129.75 &   $-$54.81 &   12.67 &   0.45 & L \\
     &  94 &     173.09 &     807.56 &   11.06 &   0.22 & R \\
     &  95 &     195.97 &     575.29 &    6.47 &   0.15 & L \\
     &  96 &     197.78 &     595.08 &    8.02 &   0.66 & R \\
     &  97 &     198.82 &     482.75 &    9.11 &   0.54 & R \\
     &     &     199.99 &     480.96 &    9.11 &   0.06 & L \\
     &  98 &     202.42 &     708.63 &    9.98 &   1.69 & R \\
     &  99 &     203.25 &     568.73 &    5.42 &   0.31 & L \\
     &     &     203.40 &     570.11 &    5.42 &   0.10 & R \\
     & 100 &     206.94 &     605.17 &    8.45 &   0.42 & R \\
     & 101 &     207.18 &     674.50 &    5.42 &   5.08 & L \\
     & 102 &     207.81 &     673.79 &    8.09 &   0.94 & R \\
     & 103 &     219.18 &     618.12 &    8.05 &   0.10 & R \\
     & 104 &     219.78 &     617.31 &    5.24 &   0.27 & L \\
     & 105 &     222.44 &     655.77 &    6.01 &   1.30 & L \\
     & 106 &     233.08 &     549.21 &    9.20 &   0.32 & R \\
     & 107 &     233.66 &     550.95 &    6.49 &   0.61 & L \\
     & 108 &     457.98 &    1295.24 &    6.30 &   0.14 & R \\
     & 109 &     465.01 &    1301.25 &    5.95 &   0.07 & L \\
     & 110 &     491.50 &    1353.89 &    6.64 &   0.13 & R \\
     & 111 &     491.93 &    1351.43 &    5.62 &   0.47 & L \\
     & 112 &     740.70 &  $-$146.15 &    6.79 &   0.35 & L \\
     & 113 &     742.94 &  $-$143.59 &    4.48 &   0.37 & L \\
     & 114 &     744.02 &  $-$128.38 &    4.00 &   0.47 & L \\
     &     &     743.76 &  $-$128.26 &    4.07 &   0.13 & R \\
     & 115 &     746.10 &  $-$142.66 &    1.02 &   0.61 & R \\
     &     &     746.14 &  $-$142.78 &    1.00 &   0.18 & L \\
     & 116 &     758.52 &   $-$90.35 &   10.17 &   0.55 & L \\
     &     &     756.98 &   $-$92.17 &   10.07 &   0.36 & R \\
     & 117 &     776.37 &   $-$92.64 &   14.01 &   0.11 & L \\
     & 118 &     800.17 &  $-$175.96 &   15.46 &   0.25 & R \\
     & 119 &     962.06 &  $-$207.57 & $-$2.53 &   0.14 & L \\
     & 120 &    1784.15 &  $-$582.29 &   15.41 &   0.21 & L \\
     & 121 &    1786.02 &  $-$584.16 &   12.94 &   0.25 & R \\
     & 122 &    1833.33 &  $-$559.66 &   14.90 &   0.12 & L \\
     & 123 &    1833.95 &  $-$559.60 &   12.56 &   0.79 & R
\enddata
\tablecomments{Coordinates are position offsets from the reference
  feature located at the origin.  Spots detected at the same location
  in RCP and LCP at nearly the same LSR velocity are treated as a
  single feature, as in \citet{fish05}.}
\end{deluxetable}

\begin{deluxetable}{rrrrrrrrrl}
\tabletypesize{\small}
\tablecaption{Zeeman Pairs\label{zeeman-table}}
\tablehead{
  \colhead{} &
  \colhead{} &
  \colhead{RCP} &
  \colhead{} &
  \colhead{} &
  \colhead{LCP} &
  \colhead{} &
  \colhead{} &
  \colhead{} &
  \colhead{} \\
  \colhead{Frequency} &
  \colhead{$\Delta$ RA} &
  \colhead{$\Delta$ Dec} &
  \colhead{v$_\mathrm{LSR}$} &
  \colhead{$\Delta$ RA} &
  \colhead{$\Delta$ Dec} &
  \colhead{v$_\mathrm{LSR}$} &
  \colhead{$B$} &
  \colhead{Separation} &
  \colhead{Pair} \\
  \colhead{(MHz)} &
  \colhead{(mas)} &
  \colhead{(mas)} &
  \colhead{(km\,s$^{-1}$)} &
  \colhead{(mas)} &
  \colhead{(mas)} &
  \colhead{(km\,s$^{-1}$)} &
  \colhead{(mG)} &
  \colhead{(mas)} &
  \colhead{Number\tablenotemark{a}}
}
\startdata
1665 &    0.00 &      0.00 & 12.36 & $-$0.24 &   $-$0.17 &    9.27 &    5.2 &  0.3 & Z1 \\
     &   99.89 &  $-$80.27 &  9.64 &   91.69 &  $-$70.85 &   13.68 & $-$6.8 & 12.5 & \\
     &  129.00 &  $-$58.36 & 14.56 &  127.59 &  $-$55.61 &   12.09 &    4.2 &  3.1 & \\
     &  193.50 &    579.87 &  9.99 &  201.09 &    568.85 &    4.01 &   10.1 & 13.4 & \\
     &  203.33 &    673.95 &  8.96 &  203.33 &    673.95 &    4.54 &    7.5 &  0.0 & \\
     &  231.83 &    581.66 & 10.15 &  230.48 &    581.26 &    5.41 &    8.0 &  1.4 & \\
     &  231.93 &    549.56 & 10.11 &  233.01 &    558.86 &    5.55 &    7.7 &  9.4 & \\
     &  277.46 &   1114.03 &  6.50 &  282.51 &   1113.27 &    4.07\tablenotemark{b} &    4.1 &  5.1 & \\
     &  301.64 &   1178.27 &  5.70 &  300.27 &   1178.27 &    3.21 &    4.2 &  1.4 & \\
     &  342.64 &   1177.95 &  5.51 &  346.17 &   1180.21 &    3.48 &    3.4 &  4.2 & \\
     &  367.61 &   1149.59 &  7.01 &  365.98 &   1147.96 &    4.54 &    4.2 &  2.3 & \\
     &  489.71 &   1348.16 &  7.21 &  488.92 &   1347.51 &    5.16 &    3.5 &  1.0 & Z3 \\
     &  548.23 &   1491.99 &  6.15 &  550.47 &   1496.78 &    3.69 &    4.2 &  5.3 & \\
     & 1785.98 & $-$596.61 & 12.16 & 1784.06 & $-$587.59 &   16.14 & $-$6.8 &  9.2 & \\
1667 &    1.10 &   $-$0.61 & 11.89 &    1.39 &   $-$0.06 &   10.09 &    5.1 &  0.6 & \\
     &  207.81 &    673.79 &  8.09 &  207.18 &    674.50 &    5.42 &    7.5 &  1.0 & Z2 \\
     &  219.18 &    618.12 &  8.05 &  219.78 &    617.31 &    5.24 &    7.9 &  1.0 & \\
     &  233.08 &    549.21 &  9.20 &  233.66 &    550.95 &    6.49 &    7.7 &  1.8 & \\
     &  457.98 &   1295.24 &  6.30 &  465.01 &   1301.25 &    5.95 &    1.0 &  9.3 & \\
     &  491.50 &   1353.89 &  6.64 &  491.93 &   1351.43 &    5.62 &    2.9 &  2.5 & \\
     &  746.10 & $-$142.66 &  1.02 &  742.94 & $-$143.59 &    4.48 & $-$9.8 &  3.3 & \\
     & 1786.02 & $-$584.16 & 12.94 & 1784.15 & $-$582.29 &   15.41 & $-$7.0 &  2.6 & \\
     & 1833.95 & $-$559.60 & 12.56 & 1833.33 & $-$559.66 &   14.90 & $-$6.6 &  0.6 & 
\enddata
\tablenotetext{a}{Pair number listed in Table 1 of \citet{slysh06}.}
\tablenotetext{b}{Alternatively, this may form part of a loose
  complete $5.6$~mG Zeeman pattern with spot 45 and another
  linearly-polarized spot.  See \citet{fish06}.}
\end{deluxetable}

\begin{deluxetable}{cccccccccccc}
\tabletypesize{\small}
\tablecaption{Magnetic Fields Inferred from Multi-Transition Overlap
  \label{brothers-table}}
\tablehead{
  \colhead{Frequency} &
  \colhead{} &
  \colhead{$\Delta$ RA} &
  \colhead{$\Delta$ Dec} &
  \colhead{$v_\mathrm{LSR}$} &
  \colhead{Frequency} &
  \colhead{} &
  \colhead{$\Delta$ RA} &
  \colhead{$\Delta$ Dec} &
  \colhead{$v_\mathrm{LSR}$} &
  \colhead{$B$} &
  \colhead{Separation} \\
  \colhead{(MHz)} &
  \colhead{Pol.} &
  \colhead{(mas)} &
  \colhead{(mas)} &
  \colhead{(km\,s$^{-1}$)} &
  \colhead{(MHz)} &
  \colhead{Pol.} &
  \colhead{(mas)} &
  \colhead{(mas)} &
  \colhead{(km\,s$^{-1}$)} &
  \colhead{(mG)} &
  \colhead{(mas)}
}
\startdata
1665 & R & 196.35 &    595.53 &    8.97 & 1667 & R & 197.78 &    595.08 &  8.02 &    8.0 & 1.5 \\
1665 & R & 200.94 &    707.45 &   10.87 & 1667 & R & 202.42 &    708.63 &  9.98 &    7.5 & 1.5 \\
1665 & R & 205.54 &    605.61 &    9.26 & 1667 & R & 206.94 &    605.17 &  8.45 &    6.9 & 1.5 \\
1665 & R & 739.60 & $-$145.53 &    2.00 & 1667 & L & 740.70 & $-$146.15 &  6.79 &$-$10.1 & 1.3 \\
1665 & L & 742.92 & $-$130.89 & $-$0.77 & 1667 & L & 744.02 & $-$128.38 &  4.00 &   40.4 & 2.7 \\
1665 & L & 774.58 &  $-$93.27 &   10.87 & 1667 & L & 776.37 &  $-$92.64 & 14.01 &   26.6 & 1.9 
\enddata
\end{deluxetable}

\clearpage

\end{document}